# Design and Evaluation of Smart-Contract-based System Operations for Permissioned Blockchain-based Systems*

Tatsuya Sato, Yosuke Himura and Jun Nemoto
Hitachi, Ltd., Research & Development Group, Center for Technology Innovation – Digital Technology

***Abstract*—Recently, enterprises have paid attention to permissioned blockchain (BC), where business transactions among inter-authorized organizations (forming a consortium) can automatically be executed on the basis of a distributed consensus protocol, and applications of BC have expanded as permissioned BC has adopted the features of the smart contract (SC), which is programmable user-defined business logic deployed in BC and executed with the consensus protocol. A single BC-based system will be built across multiple management domains (e.g., the data centers of each organization) having different operational policies (e.g., operational procedures, timing, dynamic parameters); although establishing system management and operations over BC-based systems (e.g., SC installation for updates) will be important for production uses, such multi-domain formation will trigger a problem in that executing system operations over BC-based systems will become time-consuming and costly due to the difficulty in unifying and/or adjusting operational policies, which is important in maintaining operational quality. Toward solving the problem, we propose an operations execution method for BC-based systems; the primary idea is to define operations as a smart contract so that unified and synchronized cross-organizational operations can be executed effectively by using BC-native features (e.g., consensus protocols). To be adaptable to the recent BC architecture in which participating nodes have different types of roles, we designed the proposed method as a hybrid architecture characterized with in-BC consensus establishment and execution status management and out-BC operations execution for all types of nodes operated by agents that listen to triggered events including operational instructions defined in SCs. A performance evaluation using a prototype with Hyperledger Fabric v1.2.0 shows that the proposed method can start executing operations within 3 seconds. Also, a functional evaluation indicates that the proposed method is more effective than alternatives from the aspect of cross-organizational BC-based system operations. Furthermore, a cost evaluation based on an estimation model and actual measurement shows that the total yearly cost of SC installation operations for updates at a quarterly pace for a 7-organization BC-based system could be reduced by 74 percent compared with a conventional manual method.**

## I. Introduction

Blockchain (BC) has been widely recognized as an epoch-making technology. BC in general is conceptually characterized by "decentralization," i.e., direct peer-to-peer transactions (e.g., financial transfers) that do not rely on any central third party. The origin of BC, called "public BC" (e.g., Bitcoin [2]), makes global-scale peer-to-peer transactions possible among many and unspecified participants and computing instances (called "nodes") with compute-intensive trust-preserving mechanisms (e.g., Proof-of-Work). BC is mainly characterized by consensus protocols, distributed ledgers, and hash chains, which enable it to have such a decentralized nature.

Recently, applications of BC have expanded from traditional cryptocurrencies to various forms of asset management in accordance with the recent capability of BC in dealing with "smart contracts (SCs)." Although SC is not a BC-specific term, we regard SCs in the context of BC as user-defined business logic for business contracts and transactions executed over BC. A SC for BC is shared with participating nodes, and the SC is automatically and definitely executed across the BC network.

Also, permissioned BC has been attracting significant attention in enterprise domains as an emerging technology for efficient cross-organizational business transactions. Different from public BC, which consists of many and unspecified participants, permissioned BC allows only inter-authorized organizations (forming a consortium) to construct a limited transaction scope in order to achieve a high transaction performance.

Toward realizing production uses of permissioned BC-based systems, it is necessary to establish system management and operations for BC-based systems. In general, system management is a range of work done to keep systems running stably; system operations are individual tasks done for system management. Examples of system operations include installing software, updating the version of installed software, booting/halting service processes, creating backups and restoring from them, and collecting system logs.

We foresee that a crucial problem in establishing system management for permissioned BC will arise from cross-organizational formation. We assume that even a single permissioned BC-based system will be built across multiple organizations (e.g., datacenters of each organization) having different operational policies (e.g., operational procedures, timing, dynamic parameters) with prohibited inter-organizational administrative access permissions. In such a situation, human managers from individual management domains will face high adjustment costs and a long lead time for cross-organizational operations (e.g., adjustment of operational policies including procedures, timing and dynamic parameters). Conventional operations execution methods that use manual and/or automa-

*Preliminary version was presented in BSC 2018 [1].

tion tools cannot solve the problem sufficiently. For example, sharing the same operational automation scripts among organizations and nodes is useful in unifying operational procedures, but such an approach does not have the capability of synchronized execution control including timing. As another example, the use of a job management server would be effective in timing control, but conventional job management servers are centralized and cannot be used for multi-organizational situations having prohibited inter-organizational administrative permissions.

In this paper, toward solving the problem facing such cross-organizational system management for per-missioned BC-based systems, we propose an operations execution method for BC-based systems, the primary idea behind of which is simple: defining system operations as SCs. On the basis of this, unified and synchronized cross-organizational operations can be executed effectively by using BC-native features (e.g., consensus protocols). In spite of the simplicity of the idea, however, implementing the proposed method is not easy due to the evolution of the recent architecture of permissioned BC, where participating nodes are assigned one of multiple different roles while maintaining BC features on the whole, the result being that only a limited subset of nodes can execute a SC (including that of the proposed method). That is, nodes on which operations should be executed are not used in the proposed method. We design the proposed method as applicable to this recent architecture by developing a hybrid approach in which we take advantage of BC-native features for establishing cross-domain consensus and managing execution statuses with SC-executable nodes ("in-BC" means the use of BC-native features), and we additionally introduce event-listening agent instances for executing operational instructions (defined in SC) for all nodes in corresponding responsible management domains while synchronizing with the execution statuses ("out-BC" means the use of non-BC-native instances).

The contributions of this present work are (a) an idea for realizing cross-organizational unified system operations for BC-based systems by using BC-native features such as SCs and consensus protocols, (b) the design of an execution method for applying the recent permissioned BC architecture, and (c) evaluations of the proposed method for BC-based systems done in qualitatively (functional comparison) as well as quantitatively (experimental performance measurement with prototype implementation and cost estimation-based model and actual measurement).

A preliminary paper [1] presented the design of our proposed method and initial evaluations using the operation of taking a snapshot of ledger data. In this paper, we mainly improved the evaluations by adding an operational cost evaluation based on actual measurement done using a more appropriate SC installation operation and also by adding a functional comparison.

## II. Problems

### A. Permissioned BC-based Systems

#### 1) Overview of BC Techinology

BC technology has attracted interest as the technology underlying the Bitcoin cryptocurrency. The fundamental features of BC are as follows.

- **Consensus protocol** is a kind of peer-to-peer communication protocol with which transactions among BC participating nodes are directly approved without relying on any third-party intermediary organization (e.g., consensus protocols with Proof-of-Work).
- **Distributed ledger** is replicated storage (storing approved transactions) owned by individual participants with which all participants are allowed to share and verify identical transaction data.
- **Hash chain** is a data structure used in distributed ledgers, where the hash value of a previous block (i.e., a set of transactions) is referred to by next block.

The above features make ledgers tamper-proof. On the basis of these three features, various derivative technologies have been proposed and are currently evolving.

#### 2) Smart Contracts

In the context of BC, the SC is user-defined logic for business contracts and transactions, which are automatically executed across BC participating nodes on the basis of a distributed consensus protocol. Recent popular BC platforms provide more generic SC functions that have almost the same expressive capability as Turing-complete program languages.

#### 3) Overview of permissioned BC

Permissioned BC is a BC derivative. Different from the original BC designed for many and unspecified participants, permissioned BC is not only characterized by the above fundamental features of BC and SC but is also designed for business transactions and information sharing only among inter-authorized organizations such as companies (forming a consortium). Permissioned BC allows common logic and code to be definitely executed while establishing consensus among organizations and the same results to be stored on each organization's ledger.

#### 4) Architecture of Permissioned BC-based Systems

The typical architecture of a permissioned BC-based system is shown in Fig. 1. The system consists of a (a) BC platform providing basic functions using BC technology and (b) BC applications, which are applications running on the BC platform. The BC platform not has only a consensus management function and distributed ledger, it also has an SC execution/management function that executes SCs according to requests from clients as well as a membership management function that manages BC network participants such as nodes and clients. In this paper, a request to execute/manage SCs is referred to as a "transaction (TX)." BC applications include the functions of traditional Web applications and SC implementations deployed on the BC platform. Individual participating organizations own such featured nodes and clients, cooperation among which enables business transactions across multiple organizations.

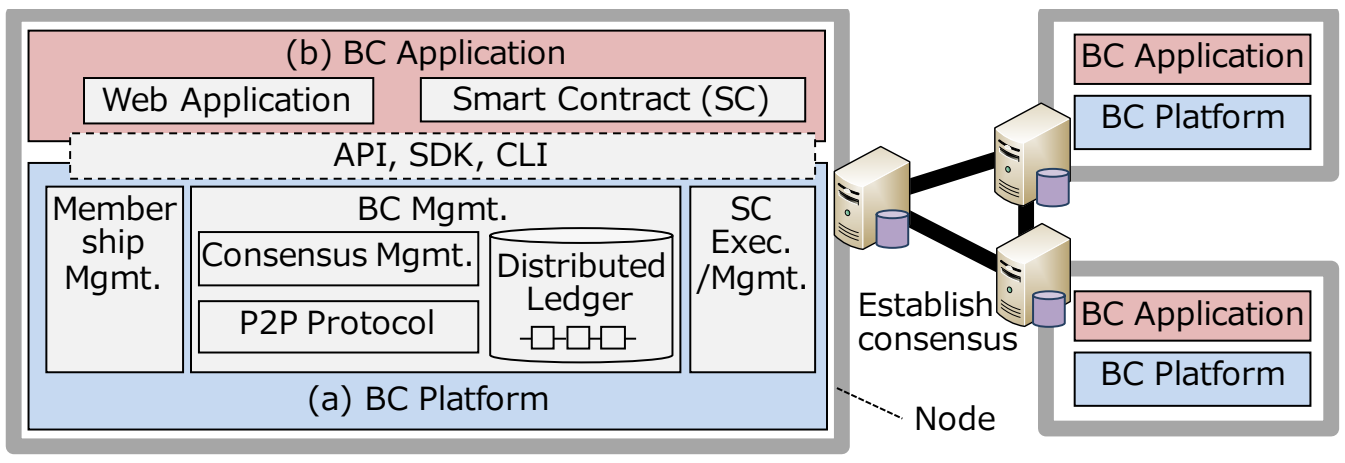

Fig. 1. Architecuture of permissioned BC systems

### B. *System Operations Execution on BC-based Systems*

In the future production phase in which permissioned BC will become widely adopted in the enterprise field, we assume that a single BC-based system will be built across multiple management domains (e.g., data centers of each organization), as shown in Fig. 2. Also, toward production uses, in general, it is necessary to maintain and improve the quality of the service of the system, and, particularly, system operations are important. Examples of system operations of permissioned BC-based systems are as follows.

- Installing a SC on each node to update or deploy it
- Booting/halting the service of the entire system.
- Taking a snapshot of ledger data and rolling back the ledger data by using the snapshot when a bug occurs in the BC and/or SC.
- Collecting system logs from each node when system problems occur and checking the status of the nodes of the entire system.
- Adding/removing nodes joining the network.

Executing these operations requires considering the conditions of both the entire and individual management domains.

In general, such system operations are executed by managers of the system by following operational policies. An operational policy defines how/when/what to execute a system operation. In this paper, an *operational policy* consists of the following three elements.

- **Operational procedure** is a step-by-step procedure for executing a system operation (e.g., a set of executed operational commands and scripts).
- **Timing** is a planned and/or on-demand time at which to start executing each operation (e.g., for periodical/emergency maintenance).
- **Dynamic parameter** is a value described in the procedure and assigned for each operation execution (e.g., command arguments such as version number for software update commands).

### C. *Problems with System Operations Execution for BC-based Systems*

System operations for BC-based systems have become difficult due to the characteristics of the situation of the multiple management domain, i.e., individual organization-specific operational policies (i.e., operational procedures, timing, and dynamic parameters) and prohibited inter-organizational access permissions.

Conventional system operation approaches include operating with a unified policy made by changing individual policies through cross-organizational adjustment (Conventional 1 and 3) and operating with individual policies (Conventional 2), but the following problems will be faced (described in Fig. 2).

- **Conventional 1.** Single manager (including a job management server) operates all nodes
  - **Problem 1.** Operations are centralized. Each manager cannot access nodes owned by other organizations because doing so would violate permissions.
- **Conventional 2.** Individual manager operates his/her own nodes
  - **Problem 2.** Lack of unified operational policy.
- **Conventional 3.** Unifying and adjusting operational policy in an offline manner before each operation by using Conventional 2
  - **Problem 3.** Time-consuming and costly due to the need to unify/adjust the operational policy conducted among human managers of individual organizations and to distribute a unified/adjusted operational policy.

Although some types of system operations (e.g., collecting logs, taking snapshots) could also be separately executed by individual organizations without a unified policy, defining cross-organizationally unified operational policies should be necessary for maintaining the quality of the system. For instance, unifying the policy for log collection (e.g., commands and log data format) will increase the visibility and interpretability of the system and resultantly will accelerate the speed of identifying route causes in the case of system problems. For another instance, unifying the policy for taking snapshots (e.g., commands and snapshot frequency) will standardize the baseline of system-wide snapshot quality by preventing there being a lack of trusted/latest snapshots for roll-back in cross-organizations. In addition, cross-organizational snapshot-taking procedures will be more solid when checking whether or not those snapshots have been tampered with through the introduction of a cross-validation step across different organizations in the procedure.

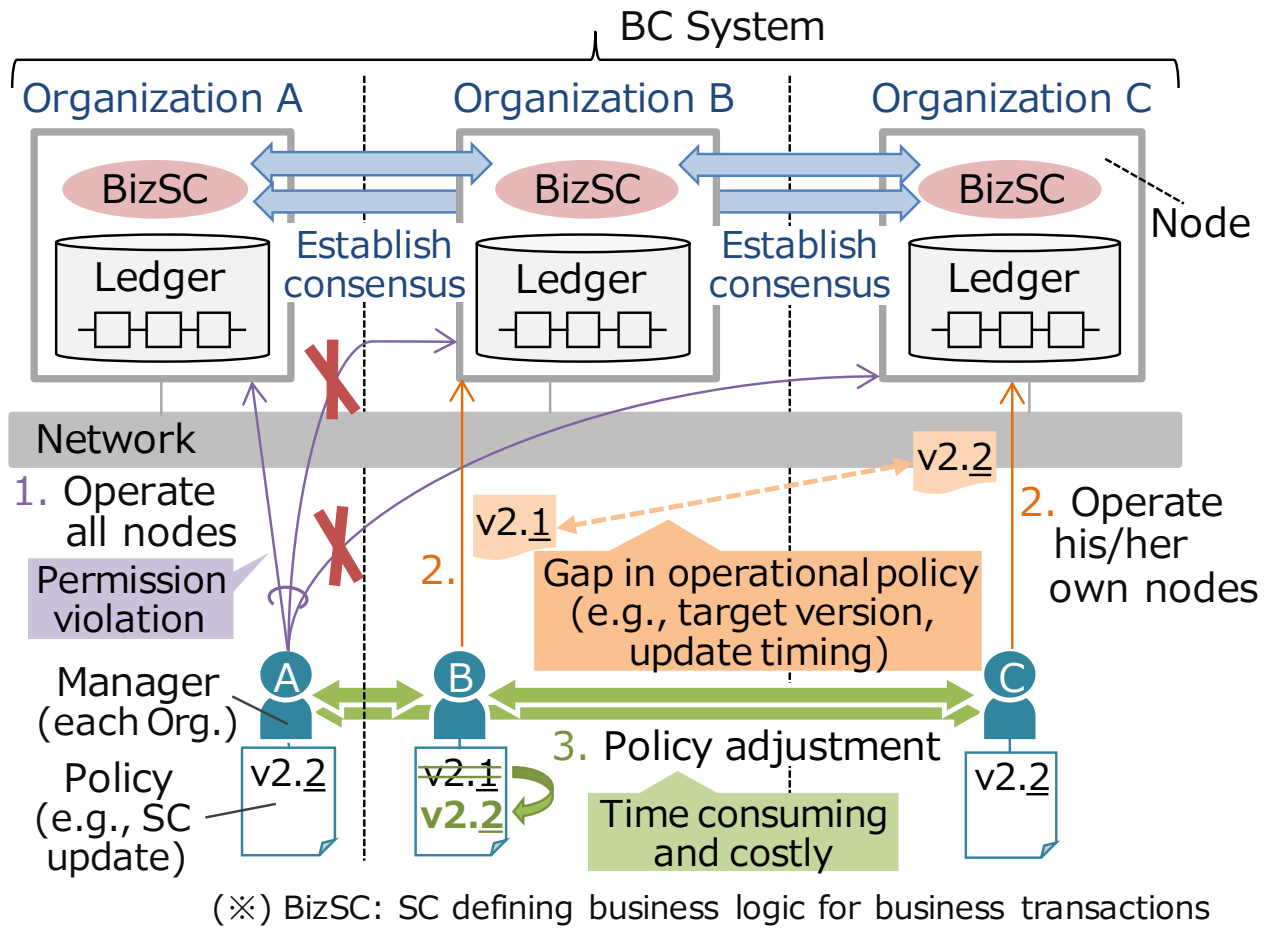

Fig. 2. Assumed system config. and problems with executing operations

## III. Methodology and Implementation

### A. Concept

Toward solving the problems, we propose an operations execution method for permissioned BC-based systems. Fig. 3 shows the concept of the proposed method. The core idea of the method is simple—"to define operations as SCs." Such SCs for system operations are hereinafter called "OpsSCs." With this approach, unified and synchronized cross-organizational operations can be executed effectively by using BC-native features such as SCs and consensus protocols. That is, instead of time-consuming and costly adjustment among human managers for each execution, BC-native consensus protocols among participating nodes will automatically execute operational procedures without there being a contradiction among the nodes following the pre-defined policies.

The proposed method makes it possible to execute synchronized system operations without inter-organizational permission violation even if the BC-based system is operated by multiple managers and built by multiple venders (this solves Problem 1). Also, by embedding operational procedures as code in SCs, the method ensures unified operational procedures with unified dynamic parameters among multiple nodes owned by different organizations (this solves Problems 2 and 3). Furthermore, an additional advantage is that the method does not require managing authentication like private keys across organizations since operational procedures are executed on the local environment of each organization. Also, the method can be realized without modifying BC platforms since it can be implemented by deploying/managing user-defined SCs.

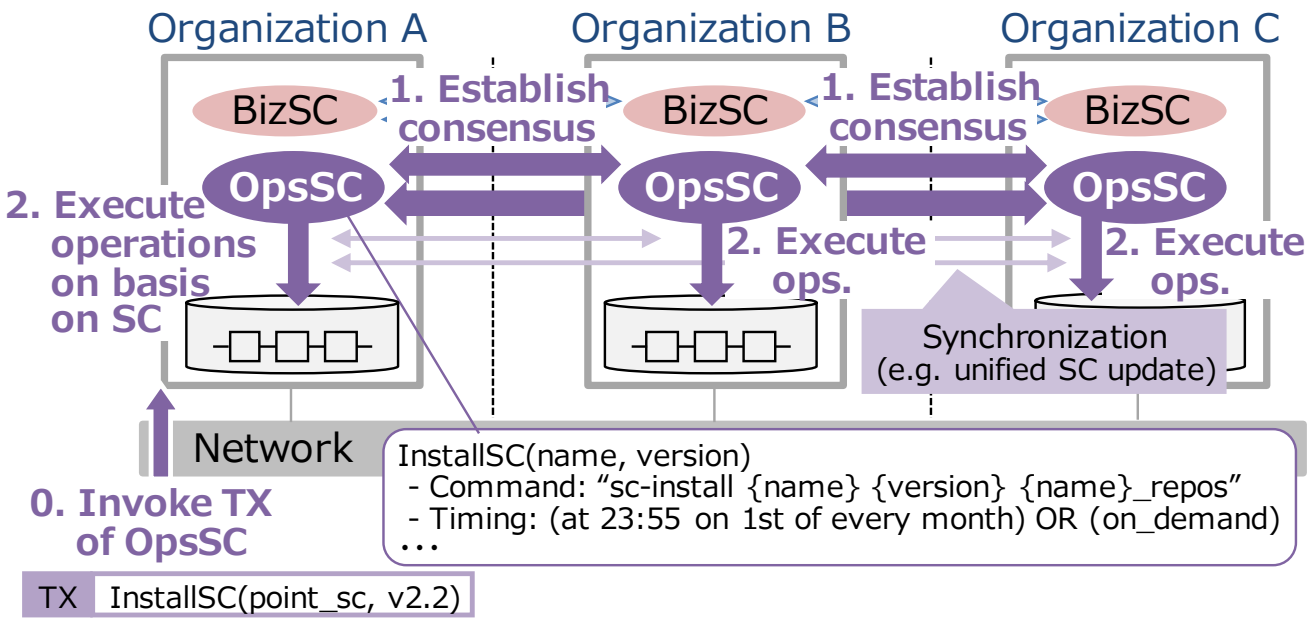

Fig. 3. Concept of proposed method

### B. Design Space

A straightforward approach to realizing our concept is to directly implement OpsSCs on a BC platform that has to be managed; this BC platform would execute general SCs that define business logic for business transactions (hereinafter BizSCs). While executing OpsSCs, this platform would execute operational instructions (e.g., system commands) written in OpsSCs.

However, realizing the straightforward approach is not easy for the following two reasons.

The first reason is that the straightforward approach performs unpreferred direct I/Os from inside SCs on resources outside SCs/BCs. The approach directly calls the local operational instructions outside SCs/BCs from inside SCs. However, such I/Os (e.g., local I/Os, calling external APIs) in SCs are not very preferable for avoiding non-deterministic TXs and layer violations in a general manner for current BC.

The second reason (the main reason) is derived from recent evolutionary trends in permissioned BC. Originally, permissioned BC was designed as an architecture in which individual participating nodes perform uniform behaviors; on the contrary, recent BC designs have been architectures in which the functionalities of the nodes are subdivided into different roles while the nodes maintain BC features on the whole and the participating nodes are assigned one of multiple different roles, the result being that the only limited nodes can execute SCs (including OpsSCs). For example, in Hyperledger Fabric [4], which is an open source enterprise-grade permissioned BC platform, nodes in the previous stable version, version 0.6, perform uniform behaviors (on the basis of the PBFT protocol [6]), whereas, in version 1.x, the roles of the nodes are subdivided; only specific nodes (called "endorsing peers") designated by a requesting client are responsible for the requested consensus and SC execution processing, and all nodes including others (called "committing peers") receive the output of SC execution delivered via an ordering service provided by special-purpose nodes (called "orderers"). As a result, the BC network of such permissioned BC is composed of nodes with different roles. Interested readers are referred to [4], [5], and [7] for details. The straightforward approach is unsuitable for the recent BC network whose nodes have different roles because an OpsSC (and the defined instructions) can be executed only on designated nodes, not necessarily on others.

To solve the limitations of operation target nodes and unpreferred direct I/Os, we introduce an agent-based hybrid design for executing OpsSCs in BC (leveraging the above features such as consensus and SC execution status management), during which operational instructions on nodes (described in OpsSC) are triggered and executed via out-BC "operational events." The nodes in BC are associated with "operational agents" listening to the execution triggers of operational instructions. In the hybrid approach, the internal processes of OpsSCs are changed. In the straightforward approach, an OpsSC executes operational instructions in an SC; in comparison, in the hybrid approach, it sends the operational instructions described in the SC as operational events to the agents. When an OpsSC issues an operational event, operational agents receive the event and then execute the embedded operational instructions on corresponding nodes. Such sharing of instructions between SCs and agents with the events helps to keep operations consistent. We believe that this hybrid design is reasonable because it can eliminate the constraints of operation targets and can also avoid the aforementioned direct I/Os inside SCs.

### C. Architecture

An architectural overview of the proposed method designed on the basis of the above agent-based hybrid approach is shown in Fig. 4. Primary components of the method are (a) a BC platform, which has to be managed, including BC applications, (b) OpsSC, which manages execution statuses and issues operational events with embedded instructions for operational pro-cedures represented in SCs (mentioned in Secs. III.A and III.B) in BC, and (c) OpsAgent, which listens to operational events issued in OpsSCs and calls operational instructions outside BCs (mentioned in Sec. III.B).

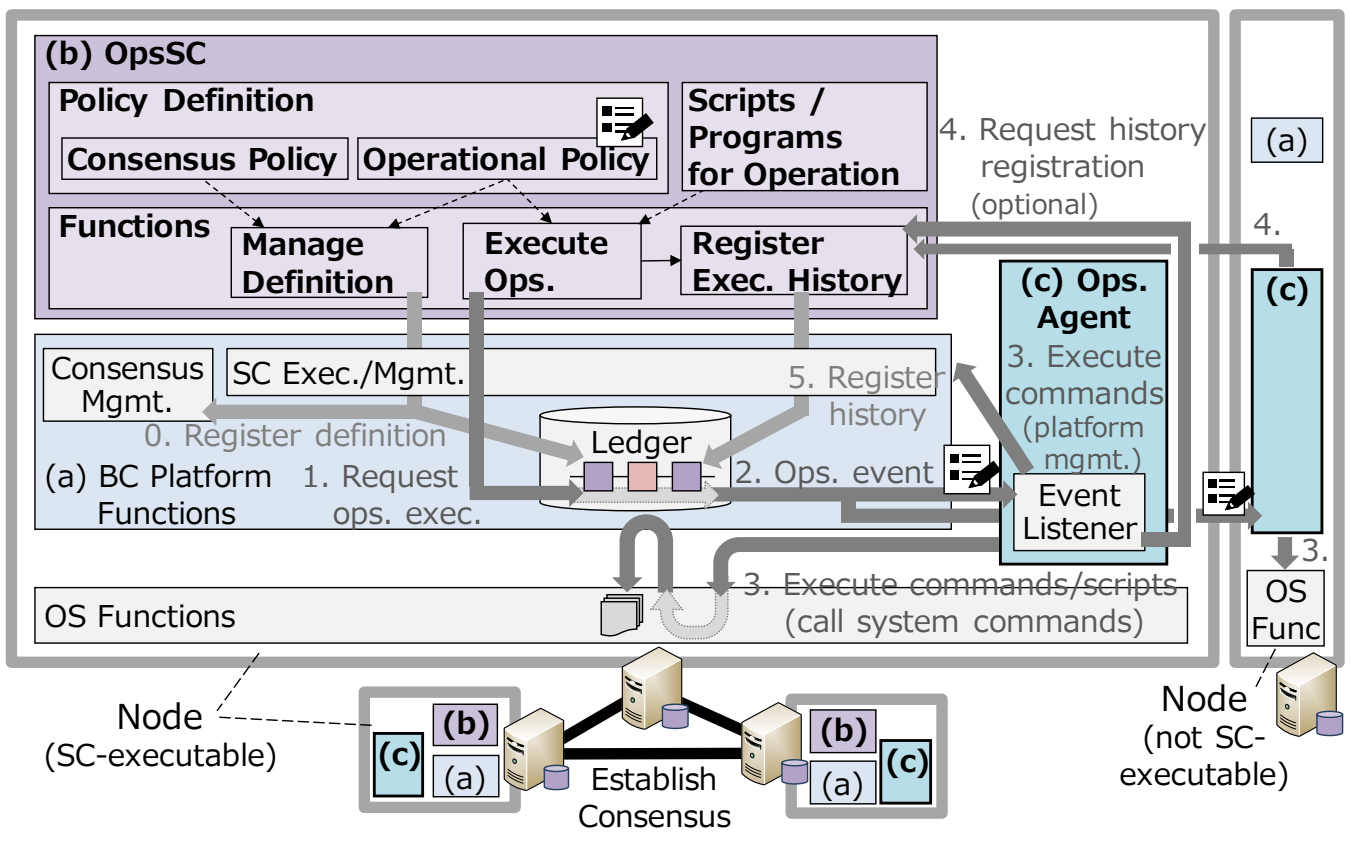

Fig. 4. BC-based system applying proposed method that introduces hybrid approach

Built on the BC platform, an OpsSC is defined by the data structures and functions of SCs. In addition to the general-purpose SC libraries provided by the BC platform, the following items are described/defined in an OpsSC as a domain-specific set for operational purposes.

- *Consensus policy* defines a policy for which nodes should participate to establish consensus. This consensus is explicitly or implicitly executed with underlying functions provided by the BC platform.
- *Operational policy* defines the operational procedures (which may be partial), parameters, and planned timing (e.g., periodicity of operations) for system operation. Operational policies are pre-described and stored in the underlying distributed ledger in advance of executing operations (e.g., pre-stored to complete negotiation among managers yet before actual execution).
- *Functions for executing operations* describe program-coded execution procedures including conditional branches for operation (using the above operational policy). These functions are coded with implementation language for SCs in the underlying BC platform.
  - Basically, the above execution procedures are described as a sequence (including conditional branches) of calls of operational commands and APIs in this OpsSC. The calls are packaged as an event that is published to agents.
  - When the operational processes are complicated, the processes also can be defined in external operation programs like scripts and can be bundled together with the OpsSC.
  - When using dynamic parameters (e.g., SC version number), the policy and/or the function defines the parameters, and the actual parameters are specified in each execution TX.
- *Function for registering execution history* records the operations histories and evidence; this process is described as part of the OpsSC. By using BC-native APIs, these histories and evidence are stored in the distributed ledger. When execution is completed and the evidence should be managed, individual agents call the function with TX embedded execution results and evidence after the event-based operations are executed on each node.

## *D. Prototype using Hyperledger Fabric v1.2.0*

On the basis of the design in Sec. III.C, we implemented a prototype of the proposed method by using Hyperledger Fabric v1.2.0, which is an open source enterprise-grade permissioned BC platform. In the current Fabric (version 1.x), whether to make the ledger tamper-proof or not depends on the consortium design (e.g., the consensus policy). If we set an appropriate design and setting for the number of organizations that participate in consensus establishment and share the distributed ledger, the ledger could be theoretically tamper-proof. Ordering services have not yet supported byzantine fault tolerant (BFT) consensus protocols, which have the ability to correctly establish consensus even in the case that there are malicious nodes. That means that until a future BFT version is supported, that version of BC leaves a single trust point. Although Fabric has not yet been perfect as mentioned above, we selected Fabric as the target for implementation and evaluation because appropriate consortium design and future BFT support will be done in the near future (actually, there are related research activities [29]); also, we anticipate that Fabric will be widely used in the enterprise field as Fabric has already been used in several production systems [5].

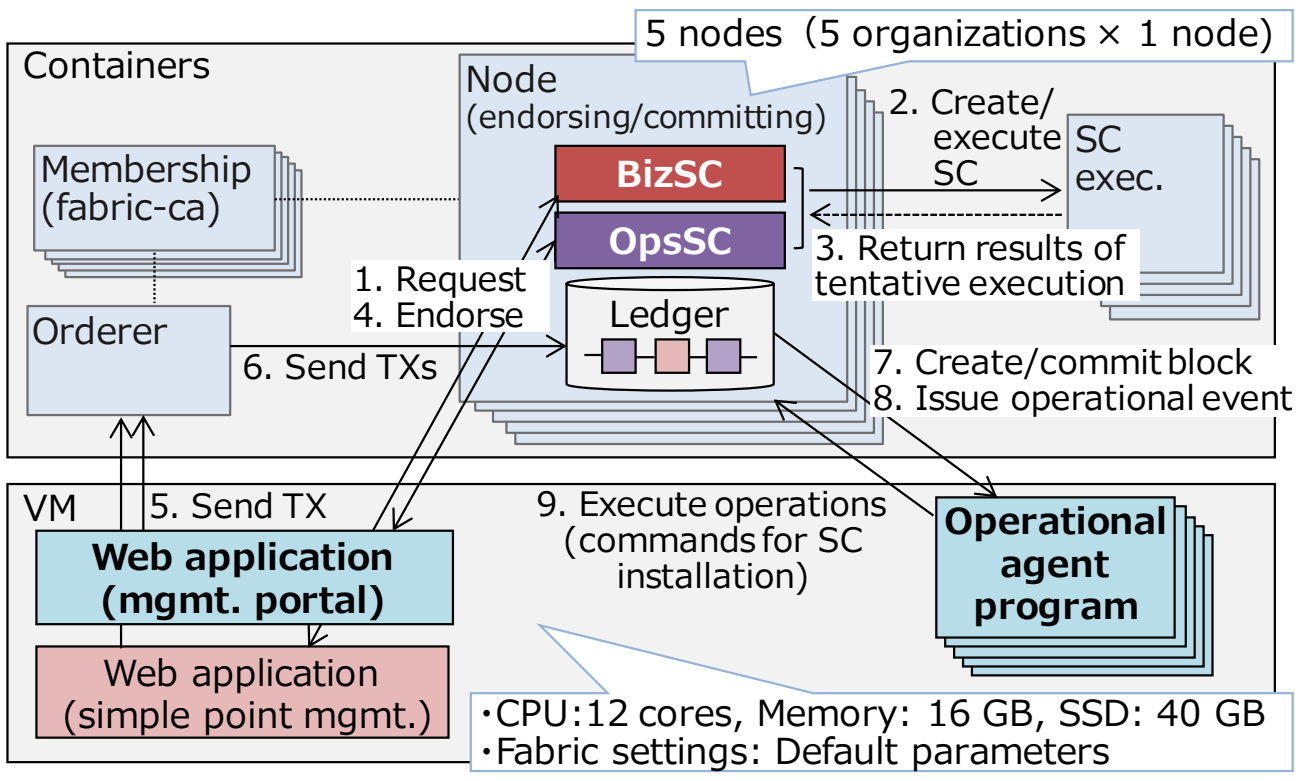

Fig. 5. Prototype using Hyperledger Fabric v1.2.0

An overview of this implementation is given in Fig. 5. For this implementation, we used Fabric v1.2.0 and a bundled sample application named “Balance-Transfer,” which had been the latest version at the time of consideration. The prototype consists mainly of a Fabric-based BC network, Web applications, and operational agent programs. The Web applications send a request for TXs of Biz/OpsSCs via Fabric’s client SDK. In addition, each operational agent program is a resident program Implemented as a client program using Fabric’s client SDK, and subscribe operational events implemented by using “Event Listener,” which is a standard function of Fabric and makes it possible to deal with custom events issued when an SC is executed. When an operational event defined as a custom event is received, each agent executes the operation commands on each node according to the command sequence embedded in the event. In the prototype, we implemented the OpsSC as a tem-plate for typical operations that execute commands sequentially. Developers can easily define a new operational policy by only inputting an operational command list into the template without SC development from scratch if the operation can be described as a simple command sequence.

## IV. Evaluations

### A. *Selecting Operation for Evaluation*

In the evaluations in Secs. IV.B and IV.D below, we select "SC installation" as a typical cross-organizational operation in Hyperledger Fabric v1.x. SC installation is required as a pre-process step for the SC deployment or update step; the SC installation step consists mainly of copying the SC source code or binary from a shared repository to all peers of each organization by using general OS commands and running the Fabric specialized command to install the SC (*peer chaincode install* command) on each peer, and after this pre-process step, we proceed with the SC deployment or update step (using *peer chaincode instantiate/upgrade* commands). This operation has to be done in a cross-organizational cooperative manner because, unless the same SC is already installed on all the organizations participating in a consensus, the SC cannot be successfully deployed/updated. For this reason, SC installation is required to be executed synchronously across organizations. It is supposed to be executed not only at planned updates but also at emergency updates, which requires agility to respond to fatal bugs that have to be fixed on demand.

The SC installation operation used in the evaluation consisted of five sequential commands as follows.

1. Cleaning up working folders on each peer.
2. Downloading a compressed specified SC source file from a shared repository to each peer.
3. Uncompressing the SC source file on each peer.
4. Installing the SC by using the "peer chaincode install" command with the SC name and version number on each peer.
5. Confirming the installed SC list on each peer by using the "peer chaincode list" command.

After these five steps for each organization, completion is reported cross-organizationally. Then, the SC deployment or update step is done.

### B. *Performance Evaluation*

#### 1) *Evaluation Method*

We conducted a performance evaluation experiment through an environmental setup with the prototype shown in Fig. 5 to verify the feasibility of the proposed method. In this experiment, each node of the prototype had an OpsSC to define the aforementioned SC installation operation. This OpsSC sequentially executed the aforementioned five commands on each peer according to the operational policy described in the OpsSC. Each node also had a simple point management application as a BizSC.

In this experiment, the following items were measured for each of the following evaluation points.

- Evaluation point: Processing time

  →Measurement item: TX execution time of OpsSC

- Evaluation point: Gap in operation timing

  →Measurement item: Difference in time to complete all operations among nodes in operation

After several trials as warm-up, TXs for the OpsSC and BizSC were respectively and sequentially invoked 1000 times. In each trial, the above items were measured.

#### 2) *Results and Considarations*

Fig. 6 shows the normalized frequency of the time the TXs took to execute the OpsSC and the gap in completion time between nodes for each trial.

**Processing time.** As shown in this figure, the TX execution time of the OpsSC, the time taken to start the operation, remained within 3 seconds; this time is faster than when human managers adjust the timing for each execution. This indicates the feasibility of the proposed method regarding processing time, although we need to be aware of a lag of a few seconds before operation starts due to the need to run the consensus protocol (which will be inevitable even if Fabric's parameters are set).

**Gap in operation timing.** The difference in the time taken to complete all operations among nodes in operation in each trial was within 300 milliseconds. This suggests that the proposed method had no serious problem with operation timing unless the required time precision for the operations is finer than the order of seconds. When finer precisions are required, additional considerations are necessary, such as setting the operation start timing in the SC.

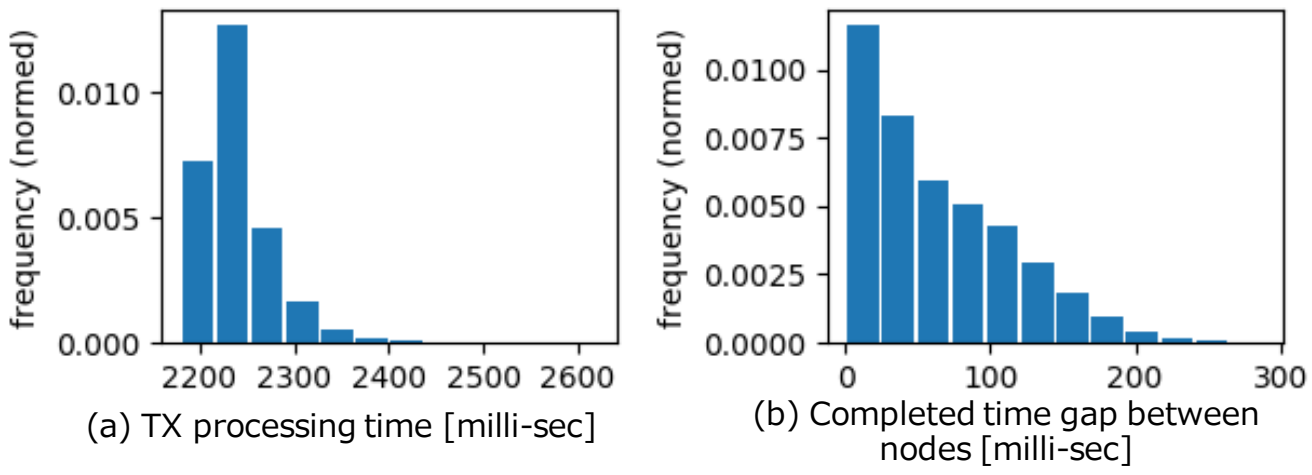

Fig. 6. Results of performance evaluation

### C. *Functional Evaluation*

#### 1) *Evaluation Method*

We validated the effectiveness of the proposed method in a functional comparison with the alternatives. First, we listed the alternatives to the proposed method. Next, we scored the achievement levels of these alternatives and the proposed method from several angles regarding cross-organizational system operations for BC-based systems. We also dealt with a comprehensive comparison based on the average score of all of the angles.

**Alternative 1 – Operations history management method using BC.** With this method, managers belonging to individual organizations collect logs of executed operations (i.e., evidence of operations history) and register the logs, which are shared with all managers, on a BC. The tamper-proof characteristics of BC make it possible to prevent registered operations histories being tampered with.

**Alternative 2 – Operations contract management method using BC.** With this method, contract documents for system operations describing agreed-upon operational standards and service levels (e.g., operations frequency and schedule) are registered and managed in BC. An advantage of this method is that if appropriate access controls are performed on these pieces of registered information, multiple organizations including business owners and IT vendors related to the BC-based system can share each other's contracts, which is referred to as an operations contract. System operations executions themselves are done manually, which is out of the scope of the method.

### 2) *Results and Considarations*

TABLE I shows the result of the functional comparison. The result of this comprehensive evaluation suggests that the proposed method is the most effective compared with the alternatives. Even if alternatives are used together, the total score would be "3.3" (the maximum value among the two methods is selected for each point), and the score is lower than the "4.7" of the proposed method. In other words, the effectiveness of the proposed method could be kept even when compared with simple combinations of alternatives.

TABLE I. FUNCTIONAL COMPARISON

| | | #0: Proposed | #1: Operations history mgmt. | #2: Operations contract mgmt. |
|---|---|---|---|---|
| Basic nature | Uniformity of operations | 5 pts. (described as code in shared SC) | 1 pt. (out of scope) | 1 pt. (out of scope) |
| Execution | Synchronization of execution timing | 5 pts. (enables cross-org. synchronized operation) | 1 pt. (out of scope) | 3 pts. (can define schedules) |
| Post-process of execution | Support for execution history management | 5 pts. [managed on BC (optional)] | 5 pts. (managed in BC) | 1 pt. (out of scope) |
| Pre-process of execution | Support for procedure-change management | 5 pts. [by updating SC, SC-based approval (optional) in BC layer] | 1 pt. (out of scope) | 5 pts. (same as proposal) |
| | Ease of procedures distribution | 5 pt. (can deploy with SCs in BC layer) | 1 pt. (out of scope) | 1 pt. (out of scope) |
| Feasibility | Ease of implementation | 3 pts. (necessary to codenize operation and contract) | 5 pts. (necessary to codenize operation history registration) | 3 pts. (necessary to codenize contract) |
| Total score (Average) | | 4.7 pts. | 2.3 pts. | 2.3 pts. |

Scoring criteria: 5 pts. … Satisfies this point, 3 pts. … Partially satisfies this point,
1 pt. … Does not satisfy this point

## D. *Cost Estimations*

### 1) *Evaluation Method*

We evaluated the effect of reducing the operational cost of system operation for BC-based systems with the proposed method compared with a conventional manual method. This evaluation was based on estimation with a cost model.

The evaluation steps are as follows.

- Step 1. Design a cost model (Sec. IV.D.1.a).
- Step 2. Experiment on setting the actual value of each unit cost for the parameters of the model (Sec. IV.D.1.b).
- Step 3. Estimate by assigning different numerical values to the remaining parameters (Sec. IV.D.2).

a) *Cost Method*

We redefined the cost model by improving the model in the preliminary paper [1]. As an evaluation metric, we first defined $C_{total}$ as the accumulated total operational cost required for executing the operations of an operational item (e.g., SC installation operation). To formally model this metric, we additionally defined $C_{plc_adj}$ as the cost of defining an operational policy including adjustment among organizations, Cops as the cost of per-operation execution, and $C_{ops_adj}$ as the adjustment cost of synchronizing timing and dynamic parameters for each execution. The cost models for the conventional and proposed methods using these metrics are shown in TABLE II. The other parameters are shown in TABLE III.

TABLE II. COST MODELS

| *(a) Conventional (Manual)* | *(b) Proposed* |
|---|---|
| $C_{total} = C_{plc_adj} + \sum_{k=1}^{n} (C_{ops_adj} + C_{ops}) \times a^{k-1} \cdots$ (a1) | $C_{total} = C_{plc_adj} + C_{dev_sc} + \sum_{k=1}^{n} C_{ops} \times a^{k-1} \cdots$ (b1) |
| $C_{plc_adj} = C_{plc_prop_unit} + (N_{org} - 1) \times C_{plc_appr_unit} \cdots$ (a2) (b2) | |
| $C_{ops_adj} = C_{ops_prop_unit} + (N_{org} - 1) \times C_{ops_appr_unit} \cdots$ (a3) | — |
| $C_{ops} = (N_{org} \times N_{node}) \times C_{exec_unit} \cdots$ (a4) | $C_{ops} = C_{trigger_unit} \cdots$ (b4) |

TABLE III. PARAMETERS OF COST MODELS

| *Parameter* | *Descriptions* | *Value* |
|---|---|---|
| $n$ | Number of executions for each operational item | 1, 2, 3, 4 |
| $N_{org}$ | Number of organizations joining BC network | 7 |
| $N_{node}$ | Number of nodes for each organization | 2 |
| $C_{plc_prop_unit}$ | Unit cost for proposer organization to define and propose policy | 79.0 MM (*) |
| $C_{plc_appr_unit}$ | Unit cost for approver organization to approve policy definition | 5.6 MM (*) |
| $C_{ops_prop_unit}$ | Unit cost for proposer organization to request to start operation per execution | 13.0 MM (*) |
| $C_{ops_appr_unit}$ | Unit cost for approver organization to approve start request per execution | 2.4 MM (*) |
| $C_{exec_unit}$ | Unit execution cost per node and execution | 6.7 MM (*) |
| $C_{trigger_unit}$ | Unit cost for triggering per execution | 0.8 MM (*) |
| $C_{dev_sc}$ | Cost of development and deployment for OpsSC | 32.9 MM (*) |
| $a$ | Improvement ratio against previous execution | 0.95 |

(*) MM: Man-minutes, values based on experiment results.

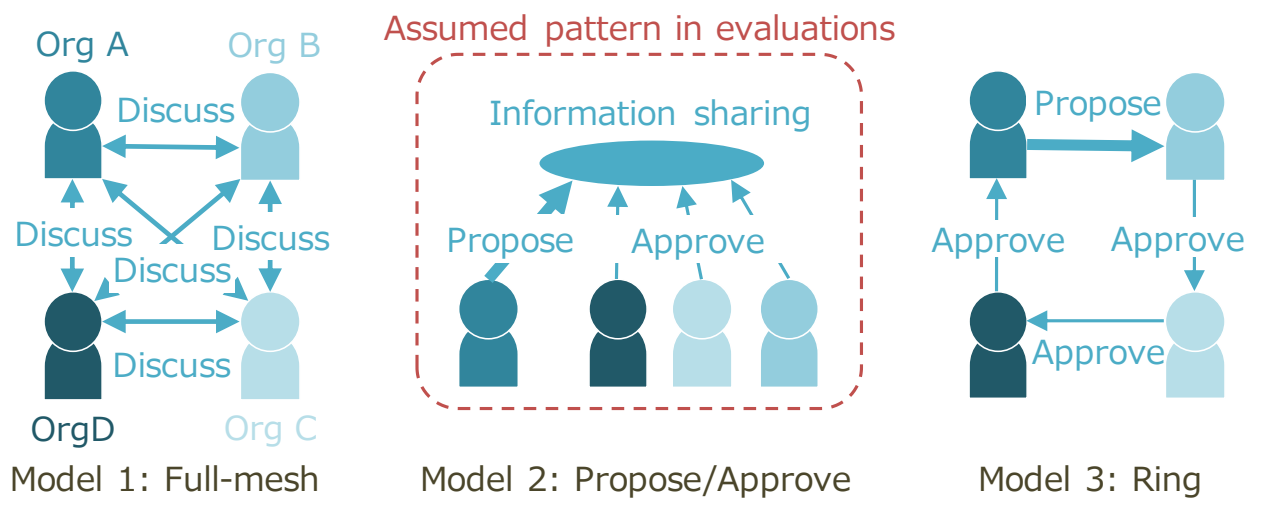

Fig. 7. Patterns of cross-organizational adjustment model

**Adjustment model.** There are several patterns for cross-organizational adjustment when defining an operational policy and/or starting to execute an operation, as illustrated in Fig. 7. In the preliminary paper [1], we used Model 1. As a result of exploring various possibilities, we created Model 2, which is a better model, for this paper. Since Model 1 is proportional to the square of the number of organizations while Model 2 is proportional to the number of one, we consider Model 2 to be more appropriate. In the method, an organization (as a proposer) first proposes the content and/or timing of an operational policy, and all other organizations (as approvers) then confirm and approve the proposal.

**Conventional manual method [TABLE II (a)].** This manual method corresponds to Conventional 3 shown in Sec. II.C. The method first defines the operational policy (mainly the procedure) while negotiating with $N_{org}$ organizations in an offline manner once; then, it executes system operations while synchronizing with the $N_{org}$ organizations in an offline manner, leading to the following assumptions. (a1) First, this method requires pre-adjustment costs $C_{plc_adj}$ once. It also requires that the adjustment for timing and dynamic parameters is conducted for each operation (i.e., $C_{ops_adj}$ + $C_{ops}$ base costs are required for each of the n executions in total), and this adjustment becomes efficient at a ratio of a due to the learning effect. (a2) First, pre-adjustment for operational procedures among organizations is performed by managers of a proposed organization and the ($N_{org}$ -1) approver organizations according to Model 2. (a3) The adjustment for synchronizing the dynamic parameters and timing for each execution is performed according to Model 2 in the same way as (a2). (a4) The operation is executed for each organization having $N_{node}$ nodes (i.e., $N_{org} \times N_{node}$ operation executions).

**Proposed method [TABLE II (b)].** (b1) The cost of adjustment for each execution becomes unnecessary because the cross-organizational procedure is shared as OpsSC code and the execution timing and dynamic parameters can be controlled (while distributed and synchronized) by the OpsSC. It instead requires additional initial costs, $C_{dev_SC}$, the cost of development and deployment for OpsSC. (b2) $C_{plc_adj}$ is the same as (a2). (b4) instead of manually executing operations on each node, this method requires that arbitrary organizations invoke the TXs of the SC to trigger the code for each execution of operation.

b) *Actual value setting based on experiment*

To verify whether system operations can be carried out according to the evaluation model shown in Sec. IV.D.1.a and to measure the actual value and ratio of each unit cost for the model, we conducted an experiment based on a role play of operation in a small system environment. We set up the system environment by using a Fabric network with five organizations, each of which had a single node. As the operational item, we targeted SC installation, as shown in Sec. IV.A.

For the manual method, five Fabric experts including the authors acted in the role of different organizations and carried out the SC installation operation manually according to the model shown in Sec. IV.D.1.a. To define the operational policy, a participant as a proposer prepared a document that defined the operational policy and sent it to all other participants as approvers by e-mail. Then, each approver confirmed the document and replied with approval by e-mail. To adjust the execution timing and share dynamic parameters, the participants used a group chat tool to propose and approve them. After adjusting in the chat, each participant executed an operation on each node according to the procedure described in the operational policy and then reported the result to the others in the chat.

For the proposed method, one of the authors developed the OpsSC and issued an SC trigger in the evaluation environment. With regard to SC development, by using the templatized OpsSC in the prototype, the author only coded to convert the command list described in the operational policy into definition information for the OpsSC. The author compiled the implemented SC, bug-fixed it while it ran in the test environment, and deployed it in the evaluation environment.

As a result of the experiment, we were able to confirm that both the conventional method and the proposed one can carry out cross-organizational operations according to the process assumed in the evaluation model. The value column with an asterisk in TABLE III shows the result of measured time for each unit cost in this experiment. The measured times executed by multiple participants are shown as mean values.

2) *Estimation Results and Considerations*

Fig. 8 shows the results of estimating the total cost $C_{total}$ of each method as a function of the number of executions *n*. As an example, actual values shown based on the measurement in the value column of TABLE III were applied to the above models. Here, man-hours were used as the unit of cost.

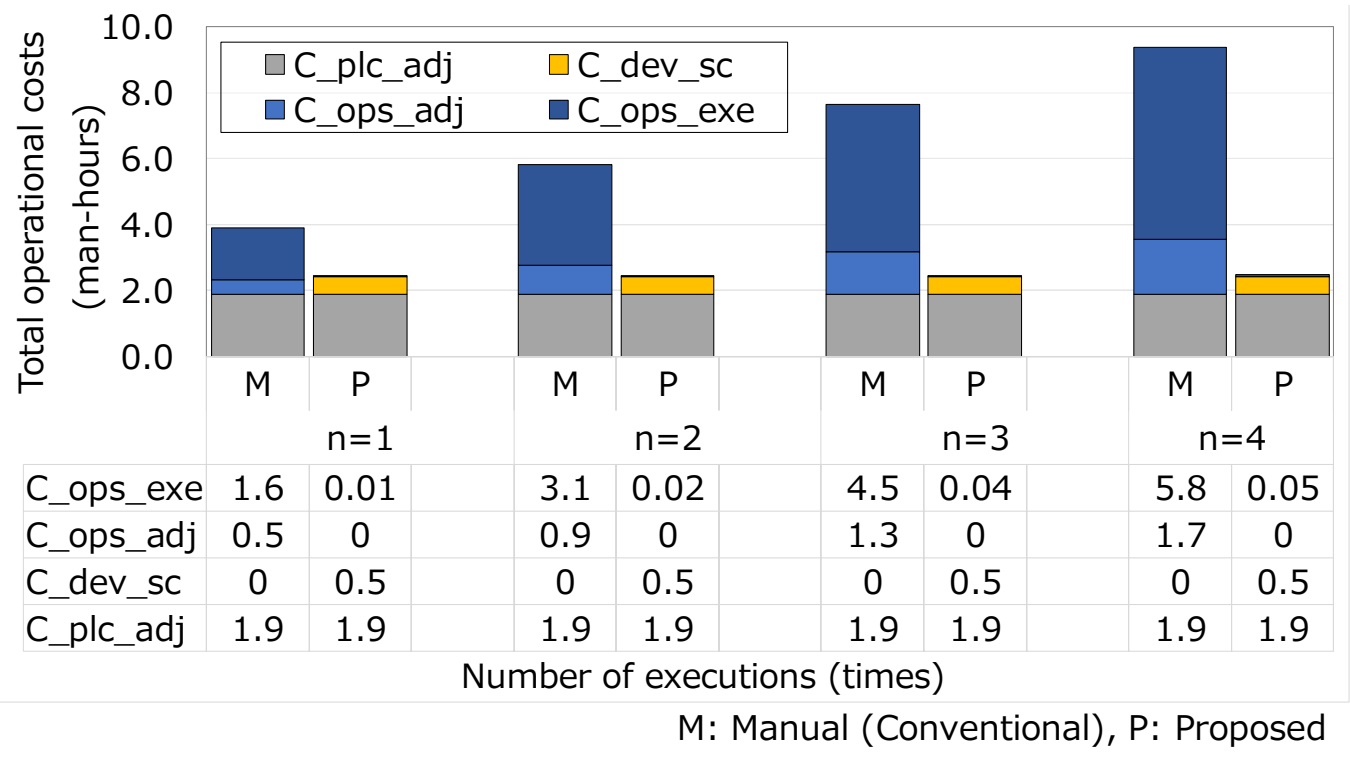

Fig. 8. Estimated results of total costs for each method

The result shows that the proposed method was not more effective at the initial stage (e.g., *n*=1) considering the initial costs; however, the total costs gradually reduced as the number of executions n increased. This means that the proposed method can greatly reduce the total costs for system operations repeatedly executed (i.e., high *n*). Furthermore, the proposed method was more effective for the BC network with more participating organizations in TABLE II. In a typical case of operations executed at a quarterly pace in a BC network in which seven organizations participate, the yearly operational cost with the proposed method was estimated as 2.5 man-hours, and that with the conventional manual method was 9.4 man-hours (74 percent reduction).

## V. DISSUCUSSIONS

### A. Other Considerations based on Actual Measurement

#### 1) Considerations assuming usages on enterprise production environment

In the experiment described in Sec. IV.D.1.b, we used a reduced system environment, so each process for performing operations was simplified. In comparison, the system operations in the production environments of actual enterprises are strict and detailed for each process, and they would require higher operational costs. The adjustment cost for each execution was lower than the expected one in the experiment because the procedure was very simple. However, actual cases may require strictly confirming the documents on the operational policy and so on and a lot of communication. Also, the cost of OpsSC development would increase because it requires scratch development if the policy and procedure become complicated.

If each process were to cost 10 times as much as the experiment in the case of an actual production environment on the basis of the assumption that the actual execution of operations would take about 1 hour, the yearly operational cost per operational item with the proposed method would be reduced by 69 man-hours compared with the manual method. Focusing on the lead time of cross-organizational operations, the time depends on the organization that becomes a bottleneck. This means that the lead time would cause a serious problem in a production environment. In the experiment, the lead time from the time the proposer notified participants of the execution timing to the time all participants finished executing operations was 11 minutes 34 seconds. If we assume that the lead time also was multiplied by 10 in the production environment, the lead time would be about 2 hours, which could be fatal, such as when a failure occurs.

#### 2) Comments from participants of experiment

From discussions with the participants after the experiment, we obtained comments on cross-organizational operations as shown below.

- With the manual method, there were worries about operational errors. In cross-organizational situations, an error made by one organization would have a big impact on the other organizations.
- Since the results of execution and evidence of them can be forged when the manual method is used, there could be trouble later. Therefore, it is useful to automatically store the evidence associated with execution in a tamper-proof ledger.
- Because there were few ways to grasp the progress of others, there was a feeling of anxiety toward the "invisibility" of the progress of others during the adjustment and execution process. It seems that this is more difficult in the case where execution synchronization with intermediate commands is required.

The issues including the comments have been solved by the current proposed method or can be solved by enhancing the method in the future.

### B. Considerations on Malicious Organizations

Our hybrid approach keeps operations consistent because operational agents execute operations according to operational instructions (embedded in operational events) defined inside a SC. This hybrid approach assumes that agents are trustworthy. Let us discuss a case in which there is a malicious agent that could eventually execute different operational instructions than those defined inside a SC.

The above malicious threat can be prevented by building a dedicated BC network for OpsSCs rather than the hybrid approach. However, this approach would require additional costs to maintain a new dedicated BC network.

Designing methods for deterring malicious threats in the hybrid approach is one direction of future research. We have considered several solutions such as a method of checking whether agents in each organization have been tampered with (e.g., checking binary and/or the processes of agents) as pre-processing for OpsSCs and a method of registering the results of executing operations (evidence) from each organization and cross-validating each piece of evidence across organizations as post-processing for OpsSCs. This pre-processing and post-processing could be also embedded within the OpsSC.

### C. Generality of Proposed Method

Besides Hyperledger Fabric used in the prototype implementation of our proposed method, various distributed ledger platforms (BC platforms in a narrow sense) have been proposed and implemented, such as Ethereum [3], Quorum [8], Corda [9], IOTA [10], Hyperledger Sawtooth [11], and Hyperledger Iroha [12]. Also, various consensus protocols for distributed ledgers including BC have been proposed and discussed (e.g., [13] and [14]). The concept and design framework proposed in our research might also be applied to these proposed platforms and protocols and might contribute to maintaining and improving the quality of systems that use these platforms and protocols. However, since the implementation method and the feasibility of OpsSC depend on the functions of the target platform and the specifications of the target protocol, it would be necessary to customize and add some functions to OpsSCs. One criterion of feasibility would be whether the functional stacks of the BC-based system as shown in Fig. 1 are similar or not. According to the current design of the proposal, at the least, the BC platform needs to be able to trigger external scripts and/or issue events in SCs to realize OpsSCs.

## VI. RELATED WORK

**Operation and management of permissioned BC-based systems.** Hyperledger Fabric [4], a permissioned BC platform, introduces a special SC called "System Chaincode (SCC)," which makes it possible to run SCs in processes and is currently used for internal processing and configuration-value sharing on the BC platform. Since our proposed method could cover the operations of an entire BC-based system including the platform, it would complement the coverage of the current SCC and contribute to maintaining the quality of a system using Fabric.

**System operations and management as code.** Tools for practicing Infrastructure as Code (IaC), such as Ansible [15], Chef [16], and Puppet [17], have been spreading to enterprise fields, which can automatically and uniformly manage and provision an IT infrastructure even on heterogeneous OSes through the abstraction and code of a domain-specific language. Furthermore, in recent years, a standard specification and related research for template-based representation for the configuration of a system running on an arbitrary cloud environment using a domain specific model have also appeared [18]. Incorporating these prior arts into our proposed method will make it possible to apply our proposed method to BC-based systems built on heterogeneous OSes and cloud environments. The method could help to extend the scope of IaC's automation to the adjustment of execution timing and dynamic parameters for each execution without cross-organizational access violations.

**Operational procedures management.** There is research on improving the efficiency of operational procedures management. [21] shows a method that extracts reusable procedure parts from documents of operational procedures to improve the efficiency of managing the procedures. [19] shows a method that discovers a history of operations by automatically analyzing raw system logs. [20] shows a method that extracts operation workflows by analyzing text-based working histories for trouble tickets. By incorporating these methods to improve the efficiency of operational procedures management into our research, cross-organizational operations could be managed and executed more efficiently (e.g., support for defining operational policies).

**Application of permissioned BC.** Various use cases and applications of permissioned BC have been be proposed and discussed, and the results are being published as articles including papers such as [27] and [28]. Considering concrete system operations by utilizing knowledge on the practical applications described in these articles would help in polishing our proposal to make it more feasible for production uses. In addition, we could also consider our proposal (especially OpsSCs and the portal application) as opening up a novel form of BC application. Our study could also contribute to the evolution of BC applications.

**Research on development of SCs.** There is research on developing SCs, such as improving SC productivity and quality. For instance, [22] reveals new research directions in BC application development, such as testing and software. [23] tries to apply traditional software design patterns to BC applications. For other instances, [24] is a study on security risk analysis for SCs, and [25] and [26] are studies on the formal verification of SCs. The results of these research fields could be also utilized to reduce security risks and improve the productivity of developing our proposed "OpsSCs."

## VII. Conclusions

In this paper, we proposed an operations execution method for permissioned blockchain (BC)-based systems. The primary idea is to define operations as smart contracts (SCs) so that unified and synchronized cross-organizational operations can be executed effectively by using BC-native features. We designed the proposed method as a hybrid architecture characterized by in-BC consensus establishment and out-BC agent-based operational instruction execution in order for it to be adaptable to the recent BC architecture in which nodes have different types of roles. A performance evaluation using a prototype implementation on Hyperledger Fabric v1.2.0 showed that the proposed method can start executing operations within 3 seconds. Furthermore, a cost evaluation using model-based estimation showed that the total cost of operations could be drastically reduced compared with a conventional manual method. In the future, we will enhance the proposed method not only for simple operational procedures but also for operational workflows.

## References

[1] T. Sato et al., "Smart-Contract based System Operations for Permissioned Blockchain," BSC 2018, p. 6, 2018.

[2] S. Nakamoto, "Bitcoin: A Peer-to-Peer Electronic Cash System," 2009.

[3] Ethereum white paper, https://github.com/ethereum/wiki/wiki/White-Paper

[4] Hyperledger Fabric, https://github.com/hyperledger/fabric.

[5] E. Androulaki et al., "Hyperledger Fabric: A Distributed Operating System for Permissioned Blockchains," arXiv:1801.10228, p. 15, 2018.

[6] M. Castro et al., "Practical Byzantine Fault Tolerance" OSDI, Vol. 99, pp. 173-186, 1999.

[7] M. Vukolić, "Rethinking Permissioned Blockchains," BCC 2017, pp. 3-7, 2017.

[8] Quorum white paper, https://github.com/jpmorganchase/quorum-docs

[9] Corda: An Introduction, https://docs.corda.net/releases/release-M7.0/_static/corda-introductory-whitepaper.pdf

[10] The Tangle, http://iotatoken.com/IOTA_Whitepaper.pdf

[11] Hyperledger Sawtooth, https://www.hyperledger.org/projects/sawtooth

[12] Hyperledger Iroha, https://www.hyperledger.org/projects/iroha

[13] M. Vukolić, "The Quest for Scalable Blockchain Fabric: Proof-of-Work vs. BFT Replication," iNetSec 2015, pp. 112-125, 2015.

[14] M. Vukolić et al., "Non-determinism in Byzantine Fault-Tolerant Replication," arXiv:1603.07351, p.20, 2016.

[15] Ansible, https://www.ansible.com/.

[16] Chef, https://www.chef.io/chef/.

[17] Puppet, https://puppet.com/.

[18] F. Glaser, Domain Model Optimized Deployment and Execution of Cloud Applications with TOSCA, SAM 2016, pp.68-83, 2016.

[19] M. Kobayashi et al., "Discovering Cloud Operation History through Log Analysis," IM 2017, pp. 959-964, 2017.

[20] A. Watanabe et al., "Workflow Extraction for Service Operation using Multiple Unstructured Trouble Tickets," NOMS 2016, pp. 652-658, 2016.

[21] H. Shikano et al., "Study on Supporting Technology for Operational Procedure Design of IT Systems in Cloud-Era Datacenters," SAC 2013, pp.405-407, 2013.

[22] S. Porru et al., "Blockchain-oriented Software Engineering: Challenges and New Directions," ICSE-C 2017, pp. 169-171, 2017.

[23] P. Zhang et al. "Applying Software Patterns to Address Interoperability in Blockchain-based Healthcare Apps," arXiv:1706.03700, p.17, 2017.

[24] E. Zhou et al., "Security Assurance for Smart Contract," BSC2018, p.5, 2018.

[25] T. Abdellatif et al., "Formal verification of smart contracts based on users and blockchain behaviors models,"BSC2018, p.5, 2018.

[26] K. Bhargavan et al., "Formal Verification of Smart Contracts," PLAS2016, pp. 91-96, 2016.

[27] K. L. Brousmiche et al., "Digitizing, Securing and Sharing Vehicles Life-cycle Over a Consortium Blockchain: Lessons Learned," p.5, BSC2018, 2018.

[28] M. Raikwar et al., "A Blockchain Framework for Insurance Processes," BSC2018, p.4, 2018.

[29] J. Sousa et al., “A Byzantine Fault-Tolerant Ordering Service for the Hyperledger Fabric Blockchain Platform,” DSN 2018, pp.51-58, 2018.